\documentclass[10pt,twocolumn,english,aps,prd,nofootinbib,preprintnumbers]{revtex4}

\usepackage[latin9]{inputenc}
\usepackage{graphicx}
\usepackage{amsmath}
\usepackage{amssymb}
\usepackage{amsfonts}
\usepackage{lscape}
\usepackage{hyperref}
\usepackage{url}
\usepackage{color}
\usepackage{bm}
\usepackage{tabularx}
\usepackage{mathtools}
\usepackage{comment}
\usepackage{soul}
\usepackage[latin9]{inputenc}

\newcommand{\be}{\begin{equation}}
\newcommand{\ee}{\end{equation}}
\newcommand{\ba}{\begin{eqnarray}}
\newcommand{\ea}{\end{eqnarray}}

\renewcommand{\[}{\begin{equation}}
\renewcommand{\]}{\end{equation}}
\def\be{\begin{equation}}
\def\ee{\end{equation}}
\def\bea{\begin{eqnarray}}
\def\eea{\end{eqnarray}}
\def\eqi{\begin{equation}}
\def\eqf{\end{equation}}
\def\eqia{\begin{eqnarray}}
\def\eqfa{\end{eqnarray}}

\definecolor{darkgreen}{rgb}{0,0.6,0}
\definecolor{darkpurple}{rgb}{0.4,0,0.6}

\makeatother

\begin{document}


\title{Confirming the uniformity of the Hubble flow with Pantheon+ Supernovae}

\author{Xiaoyun Shao$^{1}$}
\email{xiaoyun48@on.br}

\author{Carlos A. P. Bengaly$^{1}$}
\email{carlosbengaly@on.br}

\affiliation{$^{1}$Observat\'orio Nacional, 20921-400, Rio de Janeiro, RJ, Brazil}


\begin{abstract}

According to the perturbed Friedmann model, the difference between Hubble constant measurements in two rest frames, at leading order in velocity, is determined solely by the relative motion of the observers and remains unaffected by the peculiar velocities of the sources. This implies that, when averaging over a sufficiently large and distant set of sources where local nonlinear inhomogeneities are diminished, such a difference should vanish, so that the Hubble flow is statistically uniform, as predicted by the Cosmological Principle -- a core assumption of the standard cosmological paradigm. In previous works, distance measurement compilations, e.g. the CosmicFlows-3 catalogue, were used for this purpose, as it comprises a large number ($\sim 10^4$) of sources of different types. Due to the increasing amount of precise luminosity distance measurements of Type Ia Supenovae (SNe) in the last few years, in this work we investigate whether we can confirm the uniformity of the Hubble flow with low-$z$ SN distances only. By means of the Pantheon+ and SH0ES compilation, we find that the results align well with previous works based on the CF3 catalogue, and are in good agreement with the expected Hubble variance in the standard model across cosmic scales of $20-150$ Mpc. Notably, the Hubble constant difference $\Delta H_0 \approx 0$ is observed at around $85$ Mpc. Despite the smaller sample size ($\sim 10^2$ versus $\sim 10^4$) relative to CF3 at those scales, our analysis show that the Pantheon+ and SH0ES dataset supports the standard model paradigm, which indicates that the Hubble flow becomes statistically uniform at around $70-100$ Mpc, which is compatible with independent determinations of the homogeneity scale based on galaxy number counts. 

\end{abstract}

\maketitle

\section{Introduction}
\label{sec:intro}

The $\Lambda$ Cold Dark Matter ($\Lambda$CDM) model constitutes the prevailing framework in modern cosmology, describing a spatially flat Universe dominated by cold dark matter and dark energy in the form of a cosmological constant.  It successfully accounts for a broad range of observations, including the temperature anisotropies of the Cosmic Microwave Background (CMB), large-scale matter clustering probed through weak gravitational lensing, and the luminosity distance-redshift relation of Type Ia supernovae~\cite{Planck:2018vyg,Brout:2022vxf,eBOSS:2020yzd,DES:2021wwk,ACT:2023kun,Li:2023tui,di2025cosmoverse}. Nevertheless, the $\Lambda$CDM framework continues to face unresolved theoretical questions and emerging observational tensions.

The $\Lambda$CDM model, while remarkably successful, is not without its theoretical and observational shortcomings. From a theoretical standpoint, challenges such as the coincidence and fine-tuning problems raise concerns about the model's naturalness and underlying assumptions~\cite{Weinberg:2000yb}. On the observational front, a persistent tension at the $\sim 5\sigma$ level between early-Universe determinations of the Hubble constant ($H_0$) and those derived from late-time observations continues to attract significant attention~\cite[and references therein]{DiValentino:2021izs}. Additionally, recent analyses based on the first two data releases from the Dark Energy Spectroscopic Instrument (DESI) suggest potential deviations from a constant dark energy equation of state, offering possible support for dynamical dark energy models~\cite{DESI:2024mwx,DESI:2025fii}. Nonetheless, these interpretations remain contentious and continue to be rigorously examined within the scientific community~\cite{Cortes:2024lgw,Mukherjee:2024ryz,Sousa-Neto:2025gpj,colgain2025much,giovanna2024does}.

In view of these ongoing challenges, it becomes imperative to verify the validity of the foundational assumptions underlying the $\Lambda$CDM model. A central tenet of this framework is the Cosmological Principle (CP), which posits that the Universe is statistically homogeneous and isotropic on sufficiently large scales~\cite{Clarkson:2010uz,Maartens:2011yx,Clarkson:2012bg,Aluri:2022hzs}. This principle serves as the basis for the Friedmann--Lema\^{\i}tre--Robertson--Walker (FLRW) metric~\cite{Hawking:1973uf,Weinberg:2008zzc}, which underpins the standard cosmological model. However, it is important to emphasize that only statistical isotropy can be directly tested through observations. The assumption of spatial homogeneity remains indirectly inferred, as our observational access is confined to the two-dimensional intersection of our past light cone with the spatial hypersurfaces on which cosmological sources reside. Consequently, tests of homogeneity rely on the internal consistency of observables and cannot be validated through direct measurements.

Several studies suggest that the Universe becomes statistically homogeneous on scales larger than approximately $100$ Mpc. However, these estimates can only provide a consistency test of the standard model, as previously discussed, in addition to being limited by the cosmic variance of the corresponding redshift surveys used in those analyses~\cite{Hogg:2004vw, Scrimgeour:2012wt, Alonso:2014xca, Laurent:2016eqo, Ntelis:2017nrj, Goncalves:2017dzs, Goncalves:2018sxa, Goncalves:2020erb, Andrade:2022imy, Shao:2023sxk, Shao:2024qrd, Goyal:2024ctd,courtois2502search}. On scales smaller than or comparable to the statistical homogeneity scale, the Hubble flow exhibits a complex pattern of spatial variance. Within the standard FLRW paradigm, such variations are typically interpreted as arising from peculiar velocity fields, i.e., the proper motion of galaxies relative to the uniform Hubble expansion. Those bulk motions are expected to become smaller at larger scales, as the matter density perturbations decrease. The rest frame defined by the Cosmic Microwave Background serves as the standard frame for comoving observers, anchoring the first-order approximation of the Hubble flow.

Considerable observational effort has been dedicated to understanding peculiar motions in the local Universe.
Several studies of peculiar velocities~\cite{Watkins:2008hf, Feldman:2009es, Kashlinsky:2008ut, Kashlinsky:2009dw} have reported persistent bulk flows extending to unexpectedly large scales, potentially challenging the predictions of the perturbed FLRW framework underpinning the standard $\Lambda$CDM cosmology. In particular, \cite{Whitford:2023oww} found that the bulk flow amplitude at a depth of $173~\mathrm{Mpc/h}$ is $428 \pm 108~\mathrm{km/s}$, significantly exceeding the $\sim 192~\mathrm{km/s}$ predicted by $\Lambda$CDM mocks. This measurement is in excellent agreement with the independent results of \cite{Watkins:2023rll}, who employed the same methodology and utilized the most recent CosmicFlows-4 (CF4) distance catalogue. As a consequence of this effect, one would perceive a dipolar anisotropy in the $H_0$ across the sky, as examined in previous works using different approaches, e.g. comparing the $H_0$ value using sub-samples of cosmic distance catalogues in different patches of the celestial sphere~\cite{krishnan2022hints,zhai2022sample,mc2023anisotropic,stiskalek2025no}.

In light of these results, we perform an indirect test of the Cosmological Principle by examining the expansion rate $H_0$, computed as a spherically averaged quantity within concentric shells at increasing radial distances. This analysis is influenced by the observer's motion with respect to the cosmic microwave background (CMB) frame, introducing a velocity-dependent effect. According to the CP, the difference in $H_0$ values averaged over spheres in two distinct rest frames should be independent of both the peculiar velocities of the sources and the observer, as well as of the source distances -- assuming the latter are sufficiently large to suppress the impact of local bulk flows arising from nonlinear structure. Due to the significant systematic uncertainties affecting individual $H_0$ measurements, besides observational limitations e.g. the Malmqist bias, we instead consider the difference in Hubble constants evaluated in two rest frames, which is expected to be negligible at large enough scales if the Cosmological Principle actually holds true~\cite{Wiltshire:2012uh, McKay:2015nea, Bolejko:2015gmk,Kraljic:2016acj, Bengaly:2018uqp}.

Hence, our goal is to verify whether we can confirm the uniformity of the Hubble expansion utilising a single probe of the local universe -- namely, the most recent Pantheon+SH0ES Type Ia supernova (SN) compilation \cite{Brout:2022vxf}, which offers more precise distance estimates than the CosmicFlows-3 (CF3) catalogue \cite{Tully:2016ppz}. Although the size of the Pantheon+SH0ES sample is smaller than CF3 by roughly two order of magnitude, our analysis exhibits strong consistency with CF3 measurements within the framework of the standard $\Lambda$CDM model over the 20-150 Mpc distance range. 
In practice, we assess the difference of the Hubble constant $H_0$ measured in the CMB and Local Group rest frames (hereafter CRF and LRF, respectively) at different cosmic scales as a means to probe the assumption of an uniform Hubble flow from some of these scales onward -- and to see if it is consistent with the expectations from a fiducial $\Lambda$CDM cosmology. Additionally, we examine the impact of applying CRF- and LRF-like velocity boosts in random directions on the reconstructed Hubble flow, in order to perform a complimentary test of its isotropy. A significant deviation in either test could indicate a breakdown of the CP or a departure from the standard cosmological model. However, our results show no evidence of such discrepancies.

The paper is organised as follows. Section~\ref{sec:analysis} presents the observational dataset and outlines the methodological framework, including the tomographic analysis in spherical shells and the construction of rest-frame differential estimators. Section~\ref{sec:Results} details the results of our analysis, including comparisons with Monte Carlo simulations under linear $\Lambda$CDM perturbations, tests of statistical isotropy, and the impact of sky coverage. Finally, Section~\ref{sec:conclu} provides a comprehensive summary of our findings, discusses their significance in the context of the Cosmological Principle and the standard model, and highlights the prospects of forthcoming surveys.

\section{Data and method}\label{sec:analysis}

The Pantheon+SH0ES dataset represents one of the most recent and precise compilation of cosmological distances from SN observations. It encompasses 1701 updated light curves corresponding to 1550 unique SNe, spanning a redshift range from $z = 0.001$ to $2.26$ \cite{Scolnic:2021amr}\footnote{Data retrieved from the repository \url{https://github.com/PantheonPlusSH0ES/DataRelease}.}. Among these, 42 SNe Ia are located in galaxies that host Cepheid calibrators, forming the basis for the SH0ES determination of the Hubble constant. This dataset is used in both the Pantheon+ cosmological analysis \cite{Brout:2022vxf} and the SH0ES $H_0$ determination \cite{Riess:2021jrx}. The associated redshifts and peculiar velocity corrections to the SN redshifts are detailed in \cite{Carr:2021lcj}, while a comprehensive modelling of the peculiar velocity field is presented in \cite{Peterson:2021hel}. 

\begin{figure*}[htbp]
    \centering
\includegraphics[width=0.54\textwidth]{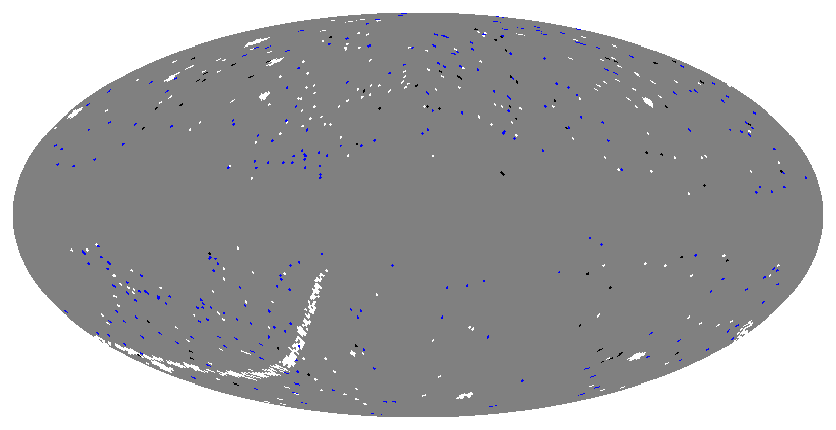} 
\caption{Mollweide projection showing the sky distribution of Pantheon+SH0ES supernovae: black represents sources with $z < 0.01$, blue corresponds to $0.01 < z < 0.03$, and white denotes $z > 0.03$.} 
\label{fig:1}
\end{figure*}

Figure~\ref{fig:1} displays the distribution of Pantheon+SH0ES supernovae across the sky, visualized using the HEALPix scheme \cite{Gorski:2004by} at a resolution of \texttt{Nside} = 64. The supernovae are colour-coded by redshift: black points correspond to nearby objects with $z < 0.01$, blue indicates the intermediate range $0.01 < z < 0.03$, and white marks the higher-redshift sample with $z > 0.03$. At higher redshifts, the non-uniformity in angular coverage becomes increasingly pronounced. We aim to test the consistency between the value of $H_0$ obtained from the observed Pantheon+SH0ES data in the redshift range $z < 0.04$, which corresponds to a scale of 150 Mpc, assuming SN redshifts at the LRF and CRF, by the same token as in previous works~\cite{Wiltshire:2012uh, McKay:2015nea, Bolejko:2015gmk, Kraljic:2016acj, Bengaly:2018uqp}. Moreover, following the discussion in the Pantheon+SH0ES analysis~\cite{Brout:2022vxf}, we adopt a minimum-redshift cut $z\ge 0.008$ to reduce potential biases from unmodelled peculiar velocities at very low redshift.

\subsection{Estimating $H_{0}$}\label{sec:est}

For redshifts in the range $z \lesssim 0.05\, h^{-1}$, the linear approximation of the Hubble law provides a reliable method for estimating the Hubble constant. In this regime, the expansion rate can be expressed as
\begin{equation}
H_0 = \frac{cz}{D(z)} + \mathcal{O}(z^2),
\label{eq1}
\end{equation}
where the distance measure $D(z)$ is identified with the luminosity distance $d_L(z)$ once we assume the SN absolute magnitude as $M_B = -19.25 \pm 0.027$ \cite{Riess:2021jrx}. This formulation isolates $H_0$ as the only cosmological parameter of relevance.

Because we restrict our analysis to distances up to roughly 150 Mpc, the higher-order redshift corrections (of order $z^2$) contribute only at the 2$\%$ level-significantly smaller than the typical $\sim$ 13$\%$  uncertainties associated with Pantheon+SH0ES distance measurements. In this low-redshift limit, the distinctions between different cosmological distance definitions (e.g., luminosity, angular diameter, comoving) are negligible. As a result, we omit the $\mathcal{O}(z^2)$ terms in the expressions that follow.

To estimate the Hubble constant, we adopt an approach based on dividing the data into concentric spherical shells, each with a radial thickness of $\Delta D$. Following the method of \cite{Bengaly:2018uqp}, we choose a broad shell width of $\Delta D = 30$ Mpc, which helps reducing the impact of nonlinear motions and allows us to probe larger distances using the Pantheon+SH0ES sample. Within each shell $s$, containing $N_s$ supernovae, we determine $H_0$ by minimizing the following chi-squared statistic 

\begin{equation}
\chi^{2}
=
\boldsymbol{\delta}^{\,\mathrm T}
\mathbf{C}_{\mathrm{stat+sys}}^{-1}
\boldsymbol{\delta},
\label{eq:cov}
\end{equation}
where the residual vector is defined as
\begin{equation}
\delta_i
\equiv
\mu_i - \mu_{\mathrm{model}}(z_i).
\end{equation}
where $\mu_i$ denotes the observed distance modulus of the $i$-th supernova, $\mathbf{C}_{\mathrm{stat+sys}}$ is the full supernova covariance matrix including both statistical and systematic contributions, and $\mu_{\mathrm{model}}$ represents the theoretical prediction, as given by the definition of the distance modulus,

\begin{equation}
\mu_{\mathrm{model}}(z)
=m_B - M_B=
5 \log_{10} \left( \frac{D(z)}{\mathrm{Mpc}} \right) + 25 ,
\end{equation}
where $D(z)$ is the luminosity distance, derived from Eq.~\ref{eq1}.
This leads to an analytic expression for the best-fit value of the Hubble constant in that shell \cite{Wiltshire:2012uh}
\begin{equation}
H_s \equiv H_0\big|_{s}
=
10^{
-\frac{1}{5}
\left(
\frac{\boldsymbol{A}^{\mathrm T}\mathbf S \mathbf B}
{\mathbf B^{\mathrm T}\mathbf S \mathbf B}
\right)
}
\, ,
\label{eq:2.3}
\end{equation}
where
\begin{equation}
\boldsymbol{A}
=
\begin{pmatrix}
\mu_1 - 5\log_{10}(c z_1) - 25 \\
\mu_2 - 5\log_{10}(c z_2) - 25 \\
\vdots \\
\mu_{N_s} - 5\log_{10}(c z_{N_s}) - 25
\end{pmatrix},
\end{equation}
$\mathbf B = (1,1,\dots,1)^{\mathrm T}$ is a column vector, and
\begin{equation}
\mathbf S = \frac{1}{2}\left(\mathbf M + \mathbf M^{\mathrm T}\right),
\qquad
\mathbf M \equiv \mathbf C_{\mathrm{stat+sys}}^{-1},
\end{equation}
where M denotes the inverse of the covariance matrix 
$\mathbf C_{\mathrm{stat+sys}}$.
$z_i$ denotes the observed redshift of source $i$. All values are obtained directly from the Pantheon+SH0ES dataset. $N_s$ is the total number of sources within the shell.

The average distance modulus of the $N_s$ sources
within shell $s$ is given by
\begin{equation}
\mu_s
=
\frac{\boldsymbol{\mu}^{\mathrm T}\mathbf{S}\mathbf{B}}
{\mathbf{B}^{\mathrm T}\mathbf{S}\mathbf{B}} \, ,
\end{equation}
where $\boldsymbol{\mu}$ denotes the column vector containing the individual distance moduli of the sources within the shell.

The corresponding average luminosity distance is then
\begin{equation}
D_s
=
10^{(\mu_s - 25)/5} \, ,
\label{eq:2.4}
\end{equation}

The associated uncertainty on the estimated Hubble constant $H_s$ takes the form

\begin{align}
\sigma_s^2
&=
\left( \nabla_{\boldsymbol{\mu}} H_s \right)^{\mathrm T}
\, \mathbf C_{\mathrm{stat+sys}} \,
\left( \nabla_{\boldsymbol{\mu}} H_s \right)
\nonumber \\
&=
\frac{(H_s \ln 10)^2}
{25\, \mathbf B^{\mathrm T} \mathbf S \mathbf B} .
\label{eq:2.5}
\end{align}

Malmquist bias, a selection effect dependent on distance but not on angular position, can still influence individual estimates of $H_s$. Rather than analyzing $H_s$ values in isolation within a given rest frame, as done in earlier works such as \cite{Wiltshire:2012uh, McKay:2015nea}, we adopt the strategy presented in \cite{Kraljic:2016acj}, which is based on the difference in $H_s$ values between two rest frames to mitigate such bias. Specifically, we focus on comparing the CMB rest-frame and the Local Group (LG) rest-frame.

The difference in Hubble parameters between frames is defined as
\begin{equation}
\Delta H_s = H_s^{\mathrm{CRF}} - H_s^{\mathrm{LRF}},
\label{eq:2.7}
\end{equation}
with an associated uncertainty
\begin{equation}
\sigma_{\Delta H_s}^2 = \left( \sigma_s^{\mathrm{CRF}} \right)^2 + \left( \sigma_s^{\mathrm{LRF}} \right)^2.
\label{eq:2.8}
\end{equation}

Since both estimates rely on the same underlying data, the uncertainties in Eq.~\ref{eq:2.8} are likely correlated. Therefore, this expression should be interpreted as an upper limit on the true error in $\Delta H_s$.

By substituting the redshifts and distances modulus transformed
to a given rest frame (RF) into Eq.~\ref{eq:2.3},
we obtain the corresponding estimator for the Hubble constant,
\begin{equation}
H_s^{\mathrm{RF}}
=
10^{-\frac{1}{5}
\frac{\left(\boldsymbol{A}^{\mathrm{RF}}\right)^{\mathrm T}
\mathbf{S}\mathbf{B}}
{\mathbf{B}^{\mathrm T}\mathbf{S}\mathbf{B}}}
\, ,
\label{eq:2.9}
\end{equation}

To gain further intuition on the expected behavior of $\Delta H_s$, we consider a simplified case in which
correlations between supernovae are neglected.
For the sake of analytical clarity, we express the $\chi^2$ statistic in terms of
luminosity distances $D_i$.
In this limit, the covariance matrix becomes diagonal,
and the $\chi^2$ function within shell $s$ reduces to \footnote{This diagonal approximation is introduced for clarity and intuition. All numerical results presented in this work are obtained using the full covariance estimator of Eq.~\ref{eq:cov}. Still, we note that we find consistent results between those estimators.}

\begin{equation}
\chi_s^2
=
\sum_{i=1}^{N_s}
\left[
\frac{D_i - (c z_i)/H_s}
{\sigma_i}
\right]^2 .
\label{eq:2.92}
\end{equation}

Minimizing this expression with respect to $H_s$
yields the corresponding estimator for the Hubble constant,
\begin{equation}
H_s^{\mathrm{RF}}
=
\frac{
\displaystyle
\sum_{i=1}^{N_s}
\frac{(c z_i^{\mathrm{RF}})^2}{\sigma_i^2}
}{
\displaystyle
\sum_{j=1}^{N_s}
\frac{
c z_j^{\mathrm{RF}}
\, D^{\mathrm{RF}}(z_j^{\mathrm{RF}})
}{
\sigma_j^2
}
}
\, .
\label{eq:2.93}
\end{equation}

The Solar System's velocity with respect to each reference frame is needed for these transformations. These are
\begin{equation}
\begin{aligned}
v^{\mathrm{CRF}} &= 369 \, \mathrm{km/s},  & (l^{\mathrm{CRF}}, b^{\mathrm{CRF}}) &= (264^\circ, 48^\circ) \, \cite{Fixsen:1996nj}, \\
v^{\mathrm{LRF}} &= 319 \, \mathrm{km/s},  & (l^{\mathrm{LRF}}, b^{\mathrm{LRF}}) &= (106^\circ, -6^\circ) \, \cite{Tully:2007ue},
\end{aligned}
\label{eq:2.10}
\end{equation}
where the coordinates $(l, b)$ denote galactic longitude and latitude, respectively.

To leading order in the dimensionless velocity $\boldsymbol{\beta}^{\mathrm{RF}} \equiv \mathbf{v}^{\mathrm{RF}}/c$, the redshift of source $i$ in a given rest frame  is related to the redshift measured in the Solar System frame via
\begin{equation}
z_i^{\mathrm{RF}} = z_i + (1 + z_i)\, \mathbf{n}_i \cdot \boldsymbol{\beta}^{\mathrm{RF}}.
\label{eq:2.11}
\end{equation}
In a general spacetime, the Doppler-induced correction to the luminosity distance takes the form \cite{Maartens:2017qoa}
\begin{equation}
D^{\mathrm{RF}}(z_i^{\mathrm{RF}}, \mathbf{n}_i) = D(z_i, \mathbf{n}_i) \left[ 1 + \mathbf{n}_i \cdot \boldsymbol{\beta}^{\mathrm{RF}} \right],
\label{eq:2.12}
\end{equation}
where $\mathbf{n}_i^{\mathrm{RF}}$ is the aberration of direction. The effect of the source's peculiar velocity $\boldsymbol{\beta}_i$ is included in the function $D(z_i, \mathbf{n}_i)$. In a linearly perturbed Friedmann model, the observed luminosity distance in the Solar System frame relates to the background value $\bar{D}(z_i)$ as \cite{sasaki1987magnitude,Bonvin:2005ps}:
\begin{equation}
\begin{aligned}
D(z_i, \mathbf{n}_i) &= \bar{D}(z_i) \left[ 1 + A_i \, \mathbf{n}_i \cdot \boldsymbol{\beta}^{\mathrm{CRF}} + (1 - A_i) \, \mathbf{n}_i \cdot \boldsymbol{\beta}_i \right]  \\
\quad 
A_i & \equiv \frac{(1 + z_i)^2}{\bar{D}(z_i) H(z_i)}.
\end{aligned}
\label{eq:2.13}
\end{equation}

Following the approach of \cite{Bengaly:2018uqp}, we can obtain the expression for $\Delta H_s$ under the assumption of a linear Hubble law for the background luminosity distance
\begin{equation}
\Delta H_s = \frac{H_0}{L_s} \sum_i \frac{c^2 z_i}{\sigma_i^2}  \boldsymbol{n}_i \cdot \left( \boldsymbol{\beta}^{\mathrm{CRF}} - \boldsymbol{\beta}^{\mathrm{LRF}} \right),
\label{eq:2.22}
\end{equation}
where $L_s \equiv \sum_i (c z_i)^2/\sigma^2_i$. Notably, contributions from the peculiar velocities $\boldsymbol{\beta}_i$ cancel out at first order in the difference between reference frames, as they are frame-independent.

Importantly, Eq.~\ref{eq:2.22} implies that if all sources within shell $s$ share the same redshift and uncertainty, the sum reduces to a directional average. For an isotropic distribution of sources over the sky, this average vanishes, resulting in $\Delta H_s = 0$. However, in realistic cases where sources have varying redshifts and uncertainties within the shell, the summation no longer cancels, and $\Delta H_s$ acquires a nonzero value -- even for an underlying isotropic distribution. This behaviour is redshift-dependent, and the general trend follows

\begin{equation}
\frac{\left|\Delta H_{s}\right|}{H_{0}} \quad\left\{\begin{array}{l}\text { grows as } z_{s} \rightarrow 0 \\ \rightarrow 0 \text { as } z_{s} \text { grows }\end{array}\right.
\label{eq:2.26}
\end{equation}

An important advantage of this analysis is that the redshifts $z_i$ are directly observed quantities, without relying on any underlying $\Lambda$CDM assumptions. At low redshift, they inherently capture anisotropies and nonlinear effects, such as local bulk flows, which are expected features of structure formation in a $\Lambda$CDM universe \cite{Kraljic:2016acj}. Although our formalism is based on a linear Doppler approximation for rest-frame transformations, it does not eliminate the nonlinear signatures encoded in the observed redshifts.

\subsection{Consistency Tests}

To assess the consistency of the Pantheon+SH0ES results with the standard $\Lambda$CDM cosmology, we perform two independent sets of 1,000 Monte Carlo (MC) simulations. Each mock catalogue preserves the angular coordinates of the original supernovae, thereby reproducing the non-uniform sky coverage inherent to the Pantheon+SH0ES dataset. The methodology for generating these simulations is outlined as follows

\begin{itemize}

\item \textbf{MC-$\Lambda$CDM:} For each supernova in the Pantheon+SH0ES sample, we generate a mock  distance modulus by sampling from the full multivariate Gaussian distribution

\begin{equation}
\boldsymbol{\mu}^{\mathrm{MC}} 
\sim 
\mathcal{N}\!\left( 
\boldsymbol{\mu}_{\mathrm{model}},\, 
\mathbf{C}_{\mathrm{stat+sys}}
\right),
\label{eq:2.27}
\end{equation}
where $\boldsymbol{\mu}_{\mathrm{model}}$ denotes the vector of predicted distance moduli from the fiducial 
$\Lambda$CDM model, and $\mathbf{C}_{\mathrm{stat+sys}}$ is the covariance 
matrix, as provided in the original data release. 


\item \textbf{MC-Boost:} In this set, we apply  CMB- and LG-like boosts to the redshift and distance modulus of each supernova. While their amplitudes are fixed to $v^{\text{CRF}}$ and $v^{\text{LRF}}$, their directions are randomised using angular coordinates $(l^{\text{MC}}, b^{\text{MC}})$. The boosted redshifts are computed as follows
    
\begin{equation}
\small
    \begin{aligned}
    cz_i^{\text{MC,CRF}} &= cz_i + 369 \left[ 
        \cos b_i \cos b^{\text{MC}} \right. \\
    &\quad \left. + \sin b_i \sin b^{\text{MC}} \cos(l_i - l^{\text{MC}}) \right], \\
    cz_i^{\text{MC,LRF}} &= cz_i + 319 \left[
        \cos b_i \cos b^{\text{MC}} \right. \\
    &\quad \left. + \sin b_i \sin b^{\text{MC}} \cos(l_i - l^{\text{MC}}) \right].
\end{aligned}
\label{eq:boosted_z}
\end{equation}

The angular coordinates defining the boost directions are drawn from uniform distributions over the full celestial sphere
\begin{equation}
b^{\text{MC}} \sim \arcsin{(-\pi/2 + \pi \mathcal{U}[0, 1])}, \quad l^{\text{MC}} \sim \mathcal{U}[0, 2\pi]. 
\label{eq:2.32}
\end{equation}

\end{itemize}

\section{Results}\label{sec:Results}

\subsection{$\Delta H_s$ from the Pantheon+SH0ES}

\begin{figure*}[htbp]
\centering
\includegraphics[width=0.54\textwidth]{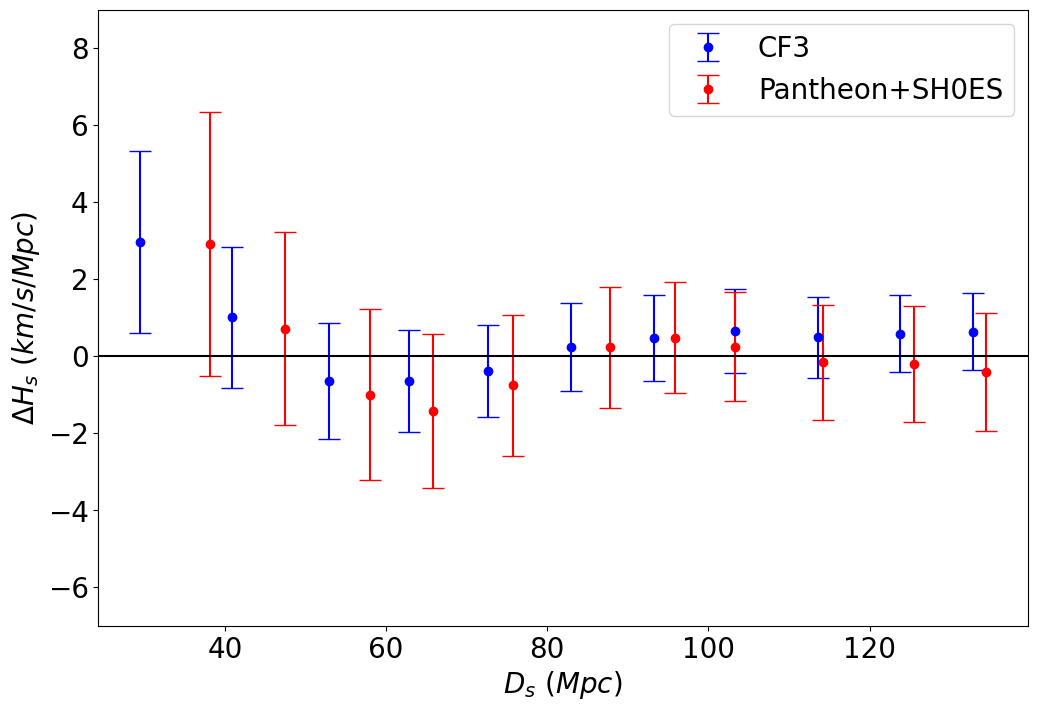} 
\caption{Difference in the average Hubble constant, $\Delta H_s$, between the CMB and Local Group  rest frames for Pantheon+SH0ES and CF3 dataset, shown as a function of shell-averaged distance $D_s$. The values are computed in successive spherical shells of width $\Delta D = 30$ Mpc, with minimum distances of $20$, $30$, and $40$ Mpc. The error bars represent 2$\sigma$ uncertainties. Numerical values are listed in Table 1.} 
\label{fig:2}
\end{figure*}

\begin{table}[ht]
\centering
\begin{tabular}{ccccc}
\hline
$D_{\min} = 20$ & $D_s$ & Shells & $\Delta H_s$ & $2\sigma_{\Delta H_s}$ \\
(Mpc) & (Mpc) & (Mpc) & (km/s/Mpc) & (km/s/Mpc) \\
\hline
& 38.19  & $20 \leq D \leq 50$   & 2.91  & 3.42 \\
& 65.79  & $50 \leq D \leq 80$   & -1.42 & 2.00 \\
& 95.84  & $80 \leq D \leq 110$  & 0.48  & 1.44 \\
& 125.46 & $110 \leq D \leq 140$ & -0.21 & 1.51 \\
\hline
$D_{\min} = 30$ & $D_s$ & Shells & $\Delta H_s$ & $2\sigma_{\Delta H_s}$ \\
(Mpc) & (Mpc) & (Mpc) & (km/s/Mpc) & (km/s/Mpc) \\
\hline
& 47.50  & $30 \leq D \leq 60$   & 0.72 & 2.50 \\
& 75.76  & $60 \leq D \leq 90$   & -0.76 & 1.84 \\
& 103.31 & $90 \leq D \leq 120$  & 0.24  & 1.41 \\
& 134.41 & $120 \leq D \leq 150$ & -0.41 & 1.53 \\
\hline
$D_{\min} = 40$ & $D_s$ & Shells & $\Delta H_s$ & $2\sigma_{\Delta H_s}$ \\
(Mpc) & (Mpc) & (Mpc) & (km/s/Mpc) & (km/s/Mpc) \\
\hline
& 58.05  & $40 \leq D \leq 70$   & -1.00 & 2.22 \\
& 87.76  & $70 \leq D \leq 100$  & 0.23 & 1.56 \\
& 114.15 & $100 \leq D \leq 130$ & -0.16 & 1.49 \\
\hline
\end{tabular}
\caption{Respectively, we present the tabulated values of the weighted average shell radius $D_s$ and the shell boundaries, along with the $\Delta H_s$ values and their associated uncertainties obtained from the Pantheon+SH0ES dataset, evaluated at $D_{\min}$ = $20$, $30$, and $40$ Mpc.}
\label{tab:combined_dHs}
\end{table}

As shown in Fig.~~\ref{fig:2}, the  difference in the Hubble constant, $\Delta H_s = H_s^{\mathrm{CRF}} - H_s^{\mathrm{LRF}}$, is plotted as a function of the weighted average shell distance $D_s$ for the Pantheon+SH0ES in red. We also plot the results obtained from the CosmicFlows-3 (CF3) dataset in blue, for the sake of comparison, taken from Table 1 of \cite{Bengaly:2018uqp}. The Pantheon+SH0ES values are computed according to the methodology detailed in Sec.~~\ref{sec:est}, using successive radial shells of width $\Delta D = 30$ Mpc. The corresponding numerical results are presented in Table~\ref{tab:combined_dHs}. 

Each point reflects the estimated $\Delta H_s$ for a given shell, with vertical error bars representing $2\sigma$ uncertainties. While Pantheon+SH0ES exhibits larger statistical errors due to its smaller sample size, both datasets show consistent trends as expected in our analytical expectation in Eq.\ref{eq:2.26}.

A notable discrepancy between the Hubble constant estimates in the CMB and LG rest frames is evident at small scales, particularly for $D_s \lesssim 40\,\mathrm{Mpc}$, where the observed values of $\Delta H_s$ are systematically positive. This result is in line with earlier studies, which indicate a more uniform in the Hubble flow in the LRF compared to the CRF \cite{Wiltshire:2012uh, McKay:2015nea}. As the shell radius increases, the magnitude of $\Delta H_s$ diminishes and gradually approaches zero, consistent with linear theory predictions and the expected emergence of large-scale statistical isotropy as the impact of local structures becomes smaller.

While the $2\sigma$ confidence intervals for $D_s > 30\,\mathrm{Mpc}$ are broad enough to include zero, the point estimates for $\Delta H_s$ are statistically indistinguishable from zero only beyond $D_s \gtrsim 85\,\mathrm{Mpc}$. This transition scale is in agreement with independent determinations of the homogeneity scale based on galaxy number counts  \cite{Hogg:2004vw, Scrimgeour:2012wt, Alonso:2014xca, Laurent:2016eqo, Ntelis:2017nrj, Goncalves:2017dzs, Goncalves:2018sxa, Goncalves:2020erb, Andrade:2022imy, Shao:2023sxk, Shao:2024qrd, Goyal:2024ctd}.

\subsection{$\Delta H_s$ from the MC realisations}

\begin{figure*}[htbp]
\centering
\includegraphics[width=0.54\textwidth]{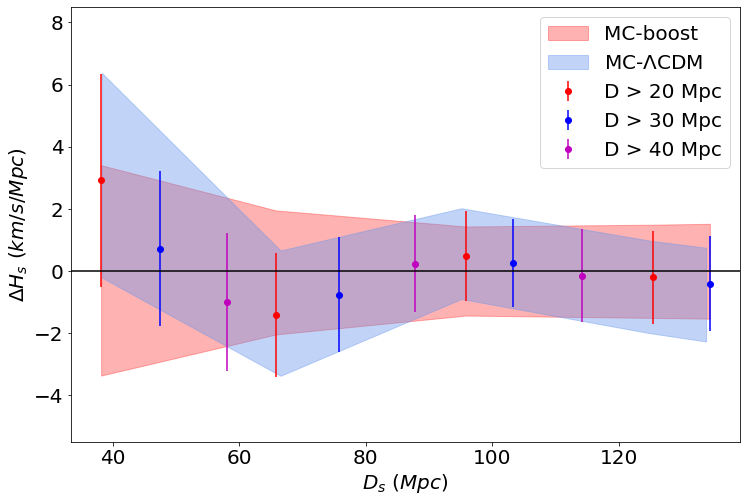} 
\caption{Comparison of the differential Hubble constant, $\Delta H_s$, between the CMB and LG rest frames based on 1,000 MC realisations. The results from observational data are contrasted with two simulated ensembles: the MC-boost realisations (shaded in light red) and the MC-$\Lambda$CDM realisations (shaded in light blue). The shaded bands represent the median $\Delta H_s$ trend across all realisations, with widths defined by the corresponding median $2\sigma_{\Delta H_s}$ uncertainties.} 
\label{fig:simu}
\end{figure*}

Fig.~\ref{fig:simu} displays the simulated $\Delta H_s$ values obtained from the MC realisations. The light blue and light red shaded regions represent the MC-$\Lambda$CDM and MC-boost ensembles, respectively. Each shaded band is centered on the median best-fit $\Delta H_s$ across all realisations, with its width corresponding to the median $2\sigma_{\Delta H_s}$ uncertainty. Observational results, previously presented in Fig.~\ref{fig:2}, are shown with $2\sigma$ error bars for three different minimum-distance thresholds: $D_{\min} > 20$ Mpc (red), $> 30$ Mpc (blue), and $> 40$ Mpc (magenta).
The results demonstrate that the $\Delta H_s$ values inferred from the MC simulations are statistically consistent with those from the observed data at all distance scales, within the $2\sigma$ confidence interval.

In the case of the MC-boost test, its concordance with respect to the observed data suggests that the CMB and LG rest frames do not exhibit statistically significant deviations from the ensemble of 1,000 randomly oriented frames, providing no compelling evidence for a violation of statistical isotropy in the local Universe

In the MC-$\Lambda$CDM test, the results also show reasonable agreement with the observational data, even on mildly nonlinear scales. This is expected, as the linear Doppler-based analysis employed here operates directly on observed redshifts, which already encapsulate the impact of local bulk motions, as detailed in Sec.~\ref{sec:est}. However, for $D_{\min} < 20$ Mpc, the limitations of the linear approximation become significant, and a fully nonlinear treatment, such as that provided by N-body simulations, would be necessary to accurately model the dynamics in this regime \cite{Kraljic:2016acj}.

\begin{figure*}[htbp]
\centering
\includegraphics[width=0.54\textwidth]{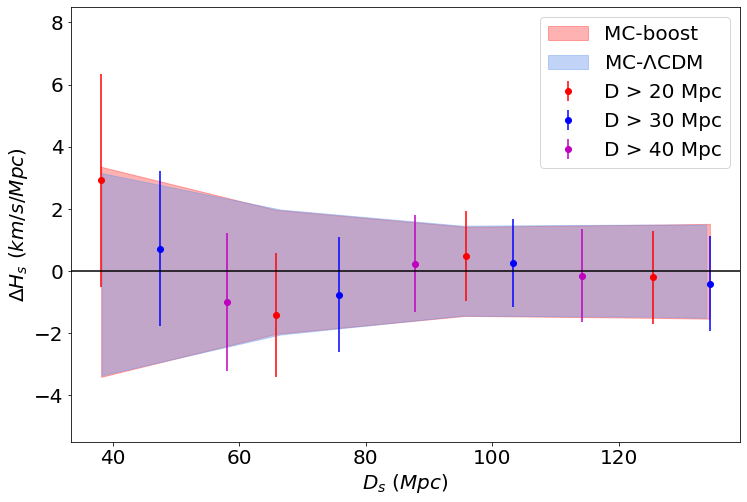} 
\caption{Same as Fig.~\ref{fig:simu}, but showing results for uniformly distributed Monte Carlo realisations.} 
\label{fig:simu_iso}
\end{figure*}

We additionally generated isotropic MC-$\Lambda$CDM and MC-boost realisations, i.e., we uniformly redistributed the source positions across the celestial sphere, instead of retaining their original angular distribution. This approach enables us to assess the impact of non-uniform sky coverage on our results. The corresponding outcomes, presented in Fig.~\ref{fig:simu_iso}, are consistent with those obtained from the Pantheon+SH0ES dataset and the original MC realisations. These findings indicate that the non-uniform angular distribution of sources in the observational data does not introduce significant bias into our analysis. 

\section{Conclusions}\label{sec:conclu}

We investigated the uniformity of the local Hubble flow by analysing the most recent and precise compilation of Type Ia Supernova cosmological distances, namely the Pantheon+SH0ES dataset. Despite its smaller sample size compared to compilation of cosmological distance measurements obtained through different methods (not only standardisable candles), as the case of the CosmicFlows-3 catalogue, the Supernova distances reported in this dataset exhibit much smaller uncertainties -- so, we expect that they should be able to capture the transition from a locally lumpy universe to a smoothed-out, statistically homogeneous one, as predicted by the cosmological principle within the standard model paradigm. 

Our analysis was carried out by fitting a linear Hubble law in tomographic radial shells within a shell width of $\Delta D_s = 30$ Mpc to reduce statistical noise, following the methodology introduced in \cite{Wiltshire:2012uh} and further developed in \cite{McKay:2015nea, Kraljic:2016acj}. Rather than examining absolute $H_0$ values in individual rest frames, we focused on the differential quantity $\Delta H_s = H_s^{\mathrm{CRF}} - H_s^{\mathrm{LRF}}$, which mitigates the impact of Malmquist bias and avoids the need to apply further corrections relative to a fiducial background value of $H_0$. According to the approach of \cite{Bengaly:2018uqp}, within the context of a linearly perturbed Friedmann--Lema\^{\i}tre--Robertson--Walker spacetime, it is expected that $\Delta H_s$ becomes negligible (even if nonzero, due to the uneven distribution of sources within a given $\Delta D_s$ shell) as $z_s$ increases. Hence, any deviation from this result would hint at a possible breakdown of the cosmological principle -- and thus the standard cosmological model -- by means of a non-uniform Hubble flow at large scales.

As illustrated in Fig.~\ref{fig:simu}, the Monte Carlo simulations constructed under the linear perturbation regime of the $\Lambda$CDM model (MC-$\Lambda$CDM) yield results that are statistically consistent with the observational data within the $2\sigma$ confidence bounds. We further verified that our findings are compatible with Monte Carlo realisations incorporating randomly oriented CMB- and LG-like boosts, which supports the absence of significant deviations from statistical isotropy. Moreover, the consistency with simulations based on uniformly distributed sources indicates that the incompleteness in the sky coverage of the Pantheon+SH0ES dataset does not introduce significant bias into our analysis.

Therefore, we are able confirm the conclusions previously reached in \cite{McKay:2015nea} and \cite{Bengaly:2018uqp}, i.e., that the Hubble flow indeed appears to become uniform at scales of $70 - 100$ Mpc -- but using Supernova observations only for the first time. Interestingly, our results agree with the homogeneity scale estimated from redshift surveys, showing that there is an interplay between them -- which helps strengthen the observational evidence for the cosmological principle. 

In the future, we expect that observational tests of the Hubble flow uniformity should be significantly improved in light of several upcoming surveys that are expected to provide  significantly enhance peculiar velocity (PV) measurements \cite{Turner:2024blz}

\begin{itemize}  

\item DESI PV will exploit spare fibres from the main DESI survey to obtain $\sim$ 133,000 Fundamental Plane (FP) and  $\sim$53,000 Tully-Fisher (TF) distances using optical spectroscopy, becoming the first large-scale survey to measure both FP and TF distances with the same instrument.

\item WALLABY, a radio survey using ASKAP, will cover 14,000 $deg^{2}$ in the southern sky over $0 < z < 0.1$, yielding tens of thousands of HI-based TF distances.

\item 4MOST Hemisphere Survey (4HS) will provide spectra for $\sim$ 6 million galaxies over 17,000 $deg^{2}$ in the southern hemisphere, including  $\sim$ 400,000 FP distances for early-type galaxies, forming the largest FP sample to date.

\item ZTF and LSST will vastly expand Type Ia supernova (SNe Ia) datasets. ZTF targets low-redshift SNe Ia ($z < 0.06$, currently  $\sim$ 3,600 events), while LSST will detect millions of SNe Ia up to $z < 0.35$, ideal for probing PVs at intermediate redshifts where SNe Ia are the only reliable tracers.

\end{itemize}

\appendix 

\section{Pantheon+ compilation without SH0ES}

\begin{figure*}
    \centering
    \includegraphics[width=0.54\linewidth]{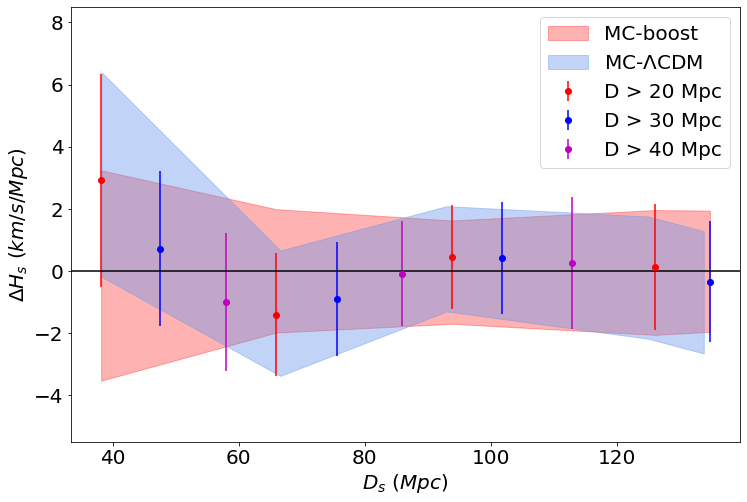}
    \caption{Same as Fig.~\ref{fig:simu}, but for Pantheon+ compilation  exclude SH0ES  instead.}
    \label{fig:noshoes}
\end{figure*}

\begin{figure*}
    \centering
    \includegraphics[width=0.54\linewidth]{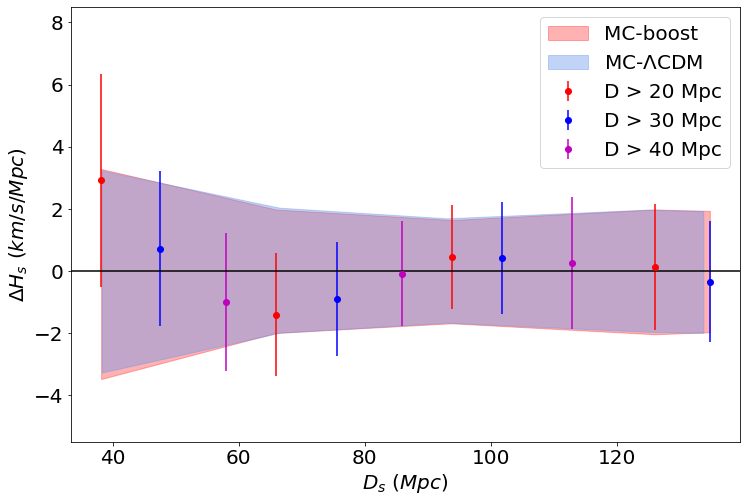}
    \caption{Same as Fig.~\ref{fig:simu_iso}, but for Pantheon+ compilation  exclude SH0ES  instead.}
    \label{fig:noshoes_iso}
\end{figure*}

For the sake of completeness, we also perform this analysis using the Pantheon+ Type Ia supernova compilation with the SH0ES subsample excluded. As shown in Fig.~\ref{fig:noshoes} and Fig.~\ref{fig:noshoes_iso}, the results remain consistent with those obtained using the Pantheon+ dataset with SH0ES included. No significant deviations are observed that would suggest a violation of the Cosmological Principle or a departure from the standard $\Lambda$CDM model.

\section{$\Delta H_s = H_s^{\mathrm{HD}} - H_s^{\mathrm{LRF}}$}

\begin{figure*}
\centering
\includegraphics[width=0.54\linewidth]{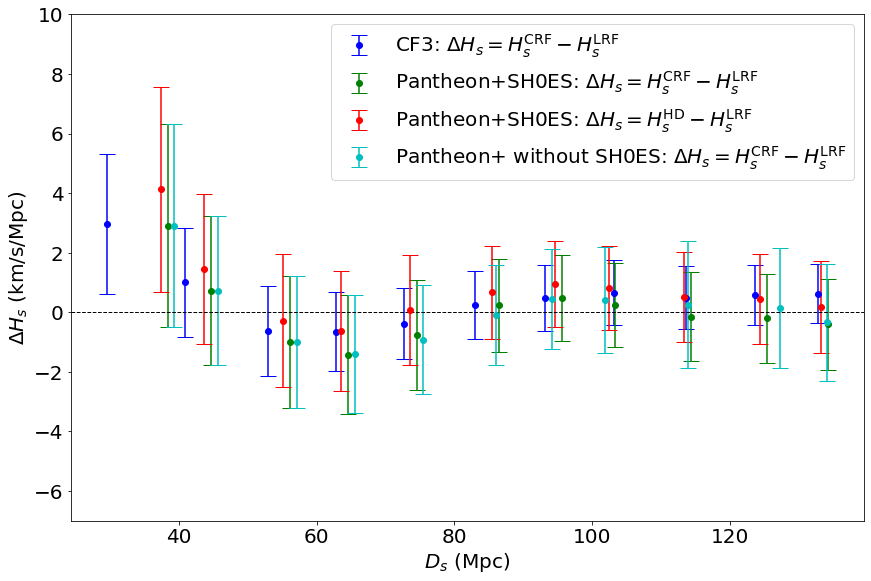}
\caption{Difference in the average Hubble constant $\Delta H_s$ as a function of shell distance $D_s$, relative to the local rest frame, using various datasets and redshift definitions. Blue points show CF3 results with $\Delta H_s = H_s^{\mathrm{CRF}} - H_s^{\mathrm{LRF}}$; green and cyan points show Pantheon+ results using $z_{\mathrm{CRF}}$, with and without SH0ES, respectively; red points use the Hubble diagram redshift $z_{\mathrm{HD}}$, with $\Delta H_s = H_s^{\mathrm{HD}} - H_s^{\mathrm{LRF}}$.}
\label{fig:com2}
\end{figure*}

As shown in Fig.~\ref{fig:com2}, we compare the differential Hubble parameter $\Delta H_s = H_s^{\mathrm{CRF}} - H_s^{\mathrm{LRF}}$ across several datasets: the Pantheon+ Type Ia supernova compilation including SH0ES (green), Pantheon+ excluding SH0ES (cyan), and the CF3 catalogue (blue). The results are in good agreement across the different samples, indicating robustness with respect to dataset selection. Additionally, we compute $\Delta H_s = H_s^{\mathrm{HD}} - H_s^{\mathrm{LRF}}$ (red), where $H_s^{\mathrm{HD}}$ is derived using the Hubble diagram redshift $z_{\mathrm{HD}}$, which is calibrated to account for both the observer's motion in the CMB frame and the peculiar velocities of the sources. Notably, the $\Delta H_s$ values based on $z_{\mathrm{HD}}$ lie closer to zero and exhibit reduced scatter, suggesting a more uniform Hubble flow compared to the results obtained using $z_{\mathrm{CMB}}$ with the SH0ES-included Pantheon+ dataset.

\section{Stability of $\Delta H_s$ Under Variations in Fiducial Cosmology}

\begin{figure*}
    \centering
    \includegraphics[width=0.54\linewidth]{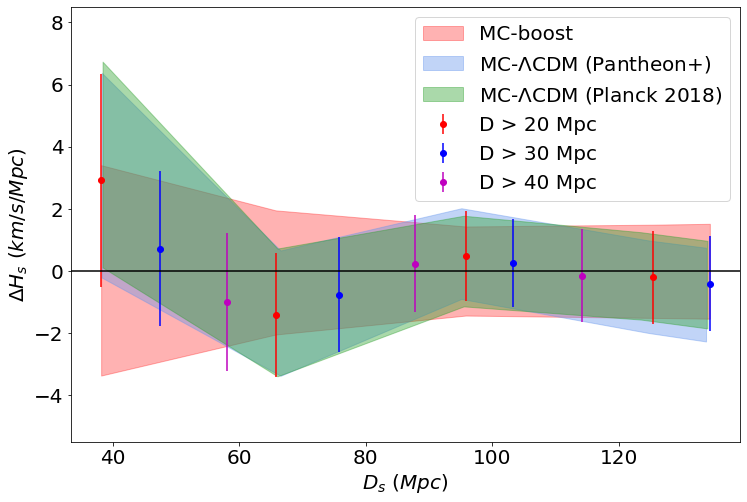}
    \caption{Same as Fig.~\ref{fig:simu}, but using a fiducial cosmology based on the Planck 2018 parameters \cite{Planck:2018vyg}.}
    \label{fig:pk2018}
\end{figure*}

\begin{figure*}
    \centering
    \includegraphics[width=0.54\linewidth]{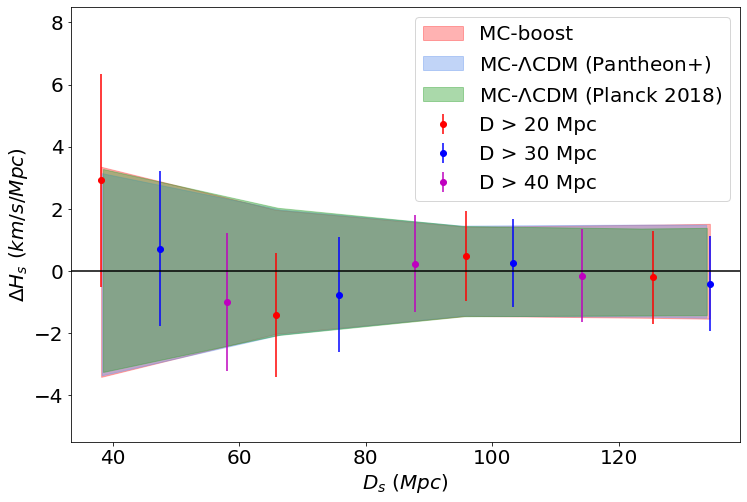}
    \caption{Same as Fig.~\ref{fig:simu_iso}, but adopting the Planck 2018 cosmological parameters.}
    \label{fig:pk2018_iso}
\end{figure*}

We have also confirmed the robustness of our results under a fiducial cosmology consistent with the Planck 2018 $\Lambda$CDM best-fitted parameters \cite{Planck:2018vyg} for the MC-$\Lambda$CDM simulations. As illustrated in Fig.~\ref{fig:pk2018} and Fig.~\ref{fig:pk2018_iso}, our conclusions remain stable under this cosmology and are not significantly affected by small variations in the underlying cosmological parameters. The results remain in good agreement with those obtained using the Pantheon+ fiducial cosmology, further supporting the consistency of our findings.

\newpage

\section*{Acknowledgements} 

XS acknowledges financial support by FAPERJ Doutorado Nota 10 fellowship, besides CAPES at the early stage of this work. CB acknowledges financial support from the CNPq grant 306630/2025-7. Some of the analysis carried out in this work made use of HEALPix.  

\bibliography{draft_v1}
\bibliographystyle{unsrt}

\end{document}